\documentclass[copyright,creativecommons]{eptcs}
\providecommand{\event}{MSFP 2016} 

\usepackage{breakurl}             
\usepackage{amsmath}
\usepackage{amssymb}
\usepackage{mathtools}

\newcommand{\Q}{\mathbb{Q}}
\newcommand{\N}{\mathbb{N}}
\renewcommand{\S}{\mathbb{S}}

\newcommand{\nlambda}{N$\lambda$}

\newcommand{\atoms}{\mathcal{A}}

\newcommand{\sembr}[1]{[\![#1]\!]}
\newcommand{\aut}{\textrm{Aut}(\atoms)}
\renewcommand{\implies}{\Rightarrow}

\title{SMT Solving for Functional Programming \\ over Infinite Structures\thanks{Supported by the Polish National Science Centre (NCN) grant 2012/07/B/ST6/01497.}}
\author{Bartek Klin \qquad \qquad Micha\l{} Szynwelski
\institute{University of Warsaw}
\email{\{klin,szynwelski\}@mimuw.edu.pl}
}
\def\titlerunning{SMT Solving for Functional Programming over Infinite Structures}
\def\authorrunning{B. Klin, M. Szynwelski}
\begin{document}
\maketitle

\begin{abstract}
We develop a simple functional programming language aimed at manipulating infinite, but first-order definable structures, such as the countably infinite clique graph or the set of all intervals with rational endpoints. Internally, such sets are represented by logical formulas that define them, and an external satisfiability modulo theories (SMT) solver is regularly run by the interpreter to check their basic properties.

The language is implemented as a Haskell module.
\end{abstract}

\section{Introduction}\label{sec:intro}

A common theme in computer science is effective manipulation of infinite but finitely presented data structures. It is one of the main features of functional programming, where computable functions, themselves infinite set-theoretic objects, are {\em bona fide} data values. In lazy programming languages such as Haskell one can also conveniently manipulate structures such as infinite lists or trees.

To achieve computability one usually restricts the interface used to manipulate infinite structures to a few basic and well-behaved operations. For example, the only way to access a function type data value is to apply it to an argument. Similarly, infinite lists provide a limited interface that allows only continuous operations on them to be implemented.

In mathematics a rich source of infinite but finitely presented objects are relational structures that are first-order definable over fixed, well understood structures. Examples include the set of ordered triples of natural numbers:
\begin{equation}\label{eq:triples}
	\big\{(a,b,c)\mid a,b,c\in\N\big\},
\end{equation}
or the infinite clique graph, with natural numbers as vertices and unordered pairs of distinct numbers as edges:
\begin{equation}\label{eq:clique}
	\big(\N,\big\{\{a,b\}\mid a,b\in\N,\ a\neq b\big\}\big).
\end{equation}
These structures are first-order defined over the set $\N$ of natural numbers with equality. On the other hand, the set of all closed intervals with rational endpoints:
\begin{equation}\label{eq:intervals}
	\big\{\{c \mid c\in\Q,\ a\leq c\leq b\}\mid a,b\in\Q\big\},
\end{equation}
or the same set partially ordered by inclusion, are defined over the set $\Q$ of rational numbers with the ordering relation $\leq$. The elements of the underlying structure, such as $\N$ or $\Q$ above, will be called {\em atoms}.

We wish to manipulate first-order definable structures effectively in the context of a functional programming language via a limited interface that can only access atoms by relations in their signature. Therefore, for example, if a set $X$ is definable over $\N$ with equality, then we do not have the ambition to check whether $X$ contains all even numbers as that property is not expressible using equality alone. On the other hand, we may wish to check whether $X$ is empty or contained in another definable set $Y$.

Computability of these and other similar conditions relies on first-order properties of the underlying structures of atoms. For example, to ensure that the set~\eqref{eq:intervals} contains some nonempty interval, one needs to know that there exist some rational numbers $a,c,b$ such that $a\leq c\leq b$. The structure of atoms should be simple enough for all such conditions to be effectively checkable. For this purpose, we shall assume that underlying structures of atoms are uniquely (up to isomorphism) determined as countable models of their first-order theories and that these first-order theories are decidable. In this paper we concentrate on two particular structures:
\begin{itemize}
\item natural numbers $\N$ with equality, understood as the unique countable model of the first-order theory of equality,
\item rational numbers $\Q$ with order $\leq$, understood as the unique countable, total, dense order without endpoints. 
\end{itemize}

Our goal is a set of programming idioms that would hide from the programmer as much as it is possible the fact that she or he is dealing with infinite sets presented by first-order formulas rather than with finite sets presented by enumerating their elements. For example, consider a program to compute the transitive closure of a binary relation. When only finite relations on a set $X$ are concerned, one can model them in Haskell as values of the type {\tt Set(a,a)}, assuming that $X$ is a set of values of a type {\tt a}. One can then code a function {\tt compose} that computes the relational composition of two relations and a function {\tt transitiveClosure} to compute the transitive closure of a relation as follows:
\begin{verbatim}
compose : (Ord a, Ord b, Ord c) => Set(a,b) -> Set(b,c) -> Set(a,c)
compose r s = sum (map (\(a,b) -> 
                          map (\(_,c) -> (a,c))
                              (filter ((==b) . fst) s))
                       r)
                         
transitiveClosure : Ord a => Set(a,a) -> Set(a,a)
transitiveClosure r = 
  let r' = union r (compose r r) 
  in if r==r' then r else (transitiveClosure r')
\end{verbatim}
using functions from the standard Haskell module {\tt Data.Set}:
\begin{verbatim}
sum = unions . elems :: Set (Set a) -> Set a
map :: Ord b => (a -> b) -> Set a -> Set b
filter :: (a -> Bool) -> Set a -> Set a
union :: Ord a => Set a -> Set a -> Set a
\end{verbatim}

One of our goals is to provide a version of the {\tt Set} type constructor that would allow the programmer to construct both finite and infinite first-order definable sets and then treat them uniformly, so that the above piece of code could be reused to compute the transitive closure of an infinite relation, internally represented by first-order formulas.

We continue the line of work started in~\cite{popl12}, where a core programming language \nlambda{} was introduced, aimed at direct manipulation of orbit-finite nominal sets~\cite{pitts-book}. These sets are typically infinite, but they can be finitely presented and they are in a strong sense equivalent to first-order definable sets over natural numbers with equality.\footnote{Other underlying structures of atoms were also considered in~\cite{popl12}, with assumptions similar to ours.} 
In~\cite{popl12}, nominal sets were constructed using so-called {\em hulls}, i.e., closures of sets under actions of automorphisms of atoms. Internally they were represented as collections of orbits.
For reasons explained in Section~\ref{sec:hulls}, we give up the orbit-based presentation of infinite sets and we use a representation based on first-order formulas instead. Technically we keep the syntax of \nlambda{} from~\cite{popl12} with few changes and semantic intuitions remain similar as well: set-typed expressions evaluate to orbit-finite sets or equivalently to first-order definable sets over atoms. However, we further propose a concrete semantics and an implementation that is significantly different from the one in~\cite{popl12}. In particular, sets are represented by first-order formulas rather than on an orbit-by-orbit basis.

Since data values are represented using logical formulas over atoms, in order to evaluate expressions one often needs to evaluate and compare such formulas to check e.g.~whether a set is empty or whether two sets are equal. This task fits in the well-researched area of {\em satisfiability modulo theories} (SMT), and there are off-the-shelf software tools tuned to that purpose. In our implementation we use the freely available Z3 checker~\cite{z3} developed by Microsoft Research, which offers satisfiability checking for first-order formulas over the theory of equality and over the theory of dense total orders without endpoints. Our implementation of~\nlambda{} intensively interacts with Z3 to analyse formulas that arise in representations of infinite data structures. We believe that this application of logical satisfiability checking in functional programming is novel; a similar application in the context of imperative programming has been developed in~\cite{lois} where mechanisms for manipulating first-order definable sets are added to the language C++.

This paper is closely related to its predecessor~\cite{popl12} and to our sister project~\cite{lois}, but the general idea of symbolic manipulation of infinite sets is far older; indeed, the entire field of constraint programming~\cite{cp} is based on it. An example of a simple programming language that integrates with an SMT solver is $\mu${Z}. The language SETL~\cite{setl} operates on set expressions, but it restricts attention to finite sets. Nominal sets, which are closely related to first-order definable sets, are manipulated in the functional programming language Fresh O'Caml~\cite{freshML}, but the main focus there is on atom binding operations, which we do not deal with here.

The structure of this paper is as follows. In Section~\ref{sec:sets-with-atoms}, we introduce first-order definable sets; the presentation is based on~\cite{KKOT15,lois,asia-thesis}. We also relate them to nominal sets~\cite{pitts-book}. In Section~\ref{sec:language}, we describe the syntax and intuitive meaning of \nlambda{} programs; this part of the paper is closely related to~\cite{popl12}, although the language is changed a little to reflect different semantic choices. In Section~\ref{sec:semantics}, a new logic-based semantics of \nlambda{} is provided. Section~\ref{sec:hulls} presents a more detailed comparison to~\cite{popl12}, and sketches an extension of the core language of Sections~\ref{sec:language}--\ref{sec:semantics} with operations to compute hulls and orbits. In Section~\ref{sec:implementation} some implementation issues are explained, and Section~\ref{sec:examples} illustrates the use of \nlambda{} on two simple examples.

A prototype implementation of \nlambda{} as a Haskell module is available for download from~\cite{nlambda}.

\medskip 

\noindent
{\bf Acknowledgments.} We are grateful to Eryk Kopczy\'nski and Szymon Toru\'nczyk, who came up with the idea of using formulas to represent orbit-finite sets with atoms, and whose work on the LOIS library for C++~\cite{lois} has been a source of constant inspiration. We also thank anonymous reviewers whose insightful comments helped us improve the paper.

\section{Sets with atoms}\label{sec:sets-with-atoms}

Fix a countably infinite relational structure $\atoms$ over some finite signature $\Sigma$. We call the elements of $\atoms$ {\em atoms}. It would be enough to assume that $\atoms$ has a decidable first-order theory and it is an {\em ultrahomogenous} structure, also known as a {\em Fra\"iss\'e limit}~\cite{hodges}. In particular, this implies that $\atoms$:
\begin{itemize}
\item is {\em $\omega$-categorical}, i.e., it is the only (up to isomorphism) countable model of its first-order theory and
\item has {\em quantifier elimination}, i.e., every first-order formula over $\atoms$ is equivalent to a quantifier-free formula. 
\end{itemize}
In this paper and for the purposes of implementation we focus on two particular structures with all these properties:
\begin{itemize}
\item $\atoms=(\N,=)$, i.e., natural numbers with equality (we call these {\em equality atoms})
\item $\atoms=(\Q,\leq)$, i.e., rational numbers with ordering (we call these {\em ordered atoms}).
\end{itemize}
For a fixed structure $\atoms$, a {\em set expression} is
\begin{itemize}
\item a variable $x$ from some fixed infinite set of atom variables, or
\item a finite sequence, written $\{\xi_1,\ldots,\xi_n\}$ (or $\{\}$ for the empty sequence), of expressions of the form
\begin{equation}\label{eq:setexp}
	\xi = e : \phi \text{ for } x_1,\ldots,x_k
\end{equation}
where $e$ is a set expression, $\phi$ is a first-order formula over $\Sigma$, and $x_1,\ldots,x_k$ are atom variables.
\end{itemize}
If $k=0$ then we write simply $e:\phi$ instead of~\eqref{eq:setexp}. We also omit $\phi$ if it is the always true formula $\top$.

The set of free variables in a set expression is defined inductively by:
\begin{align*}
	FV(x) &= \{x\} \\
	FV(e : \phi \text{ for } x_1,\ldots,x_k) &= FV(e) \cup FV(\phi) \setminus \{x_1,\ldots,x_k\} \\
	FV(\{\xi_1,\ldots,\xi_n\}) &= FV(\xi_1)\cup\cdots\cup FV(\xi_n)
\end{align*}
where $FV(\phi)$ is the standard set of free (atom) variables in a first-order formula. 
A {\em valuation} for a set expression $e$ is a function $v:FV(e)\to\atoms$. A set expression $e$ together with a valuation $v$ denotes a set (or an atom) $\sembr{e}_v$ in the expected way:
\begin{align*}
	\sembr{x}_v &= v(x) \\
	\sembr{e : \phi \text{ for } x_1,\ldots,x_k}_v &= \left\{\sembr{e}_{v[x_i\mapsto a_i]} \mid a_1,\ldots,a_n\in\atoms \text{ s.t. } \atoms,v[x_i\mapsto a_i]\models\phi\right\} \\
	\sembr{\{\xi_1,\ldots,\xi_n\}}_v &= \sembr{\xi_1}_v\cup\cdots\cup\sembr{\xi_n}_v
\end{align*}
where $\atoms,v\models \phi$ means that the formula $\phi$ holds in $\atoms$ with the valuation $v$ of the free variables in $\phi$. We say that a set of the form $\sembr{e}_v$ is {\em definable} over $\atoms$.

Standard set-theoretic tricks can be used to encode ordered pairs (e.g. as Kuratowski pairs $(x,y)=\{\{x\},\{x,y\}\}$), tuples (as nested pairs), and integers (e.g. as von Neumann numerals $n=\{0,1,\ldots,n-1\}$). 

For example, over equality atoms, the expression
\[
	\{(x,y) : \neg(x=y) \text{ for } x,y\},
\]
with the empty valuation denotes the set of ordered pairs of distinct atoms. The same definition works for ordered atoms, where we see $x=y$ as shorthand for $x\leq y \land y\leq x$. Over ordered atoms, the expression
\[
	\{x : x\leq u \text{ for } x,\ y : w\leq y \text { for } y \}
\]
with a valuation $u\mapsto 2,\ w\mapsto 5$, denotes the set of all atoms outside of the open interval $(2;5)$. The same set is denoted by the expression
\[
	\{x : x\leq u\lor w\leq x \text{ for } x\}
\]
with the same valuation.

We shall restrict attention to {\em well-typed} expressions with a set of types defined by:
\begin{equation}\label{eq:sets-types}
	\tau,\rho ::= \mathsf{A} \mid \mathsf{N} \mid (\tau,\rho) \mid \S\tau
\end{equation}
where $\S$ is a unary type constructor, with $\S\tau$ meant to be the type of sets whose elements are of type $\tau$. Set expressions are provided with types by the following relation (actually, a partial function):
\[
	\dfrac{}{x:\mathsf{A}} \qquad
	\dfrac{}{n:\mathsf{N}} \qquad
	\dfrac{e_1:\tau \quad e_2:\rho}{(e_1,e_2):(\tau,\rho)} \qquad
	\dfrac{e_1:\tau \quad \cdots \quad e_n:\tau}
	         {\{e_1:\phi_1\cdots,\ldots,e_n:\phi_n\cdots\} : \S\tau}
\]
where $x$ ranges over atom variables and $n$ over integers. Essentially it is required that all elements of a well-typed set have the same type. Pairs and integers are treated separetely here, since neither Kuratowski pairs nor von Neumann numerals are well typed in this sense.

The above constructions appear in the literature under various guises. Indeed, sets definable over atoms $\atoms$ are essentially {\em first-order interpretable} structures over $\atoms$ in the sense of model theory~\cite{hodges}. They also correspond to nominal sets~\cite{pitts-book}; we sketch this connection briefly as it relates this paper to previous work~\cite{popl12} on extending functional programming to sets with atoms.

Consider a set $X$ with a group action $\_\cdot\_:\aut\times X\to X$ of the automorphism group of the structure $\atoms$. A set $S\subseteq\atoms$ {\em supports} an element $x\in X$ if $\pi\cdot x=x$ for every $\pi\in\aut$ such that $\pi(a)=a$ for all $a\in S$. If every element of $X$ has some finite support, then $X$ is called {\em $\atoms$-nominal}. For $\atoms$ equality atoms this specializes to the notion considered in~\cite{pitts-book}. 

A function $f:X\to Y$ between nominal sets is {\em equivariant} if $f(\pi\cdot x)=\pi\cdot f(x)$ for every $x\in X$ and $\pi\in\aut$.

An {\em orbit} of an element $x\in X$ is the set $\{\pi\cdot x \mid \pi\in\aut\}\subseteq X$. Orbits form a partition of the $\atoms$-nominal set $X$; we call $X$ {\em orbit-finite} if it has finitely many orbits.

For every set expression $e$ without free variables, the set $\sembr{e}_\emptyset$ is equipped with a canonical group action of $\aut$: for $e' : \phi \text{ for } x_1,\ldots,x_k$ a part of $e$, and for $a_1,\ldots,a_n\in\atoms$ such that $\atoms,[x_i\mapsto a_i]\models\phi$, define
\[
	\pi\cdot\sembr{e'}_{[x_i\mapsto a_i]} = \sembr{e'}_{[x_i\mapsto \pi(a_i)]};
\]
$\atoms,[x_i\mapsto \pi(a_i)]\models\phi$ follows from $\pi$ being an automorphism of $\atoms$, since $FV(\phi)\subseteq\{x_1,\ldots,x_n\}$. It is easy to see that $\sembr{e'}_{[x_i\mapsto a_i]}$ is supported by $\{a_1,\ldots,a_n\}$, so $\sembr{e}_{\emptyset}$ is a $\atoms$-nominal set. Moreover, the set is orbit-finite; this follows from the fact that for every $\omega$-categorical structure $\atoms$, the set $\atoms^n$ with the pointwise action of $\aut$ is orbit-finite, by the celebrated Ryll-Nardzewski theorem from model theory~\cite{hodges}.

This means that every set definable by an expression without free variables is $\atoms$-nominal and orbit-finite. The converse also holds: every $\atoms$-nominal, orbit-finite set is equivariantly bijective to a set of the form $\sembr{e}_\emptyset$ for some set expression $e$. Moreover, if pairs are included in the language of expressions, one can choose $e$ to be well-typed. Details of this correspondence are developed in the first chapter of~\cite{asia-thesis}.

In~\cite{popl12}, a functional programming language \nlambda{} was designed to compute and manipulate orbit-finite nominal sets. There, infinite structures were internally represented on an orbit-by-orbit basis using a representation theorem from~\cite{lmcs14} saying that every single-orbit set is in equivariant bijection with a set of tuples of atoms quotiented by an equivalence relation of a certain shape. In this paper we continue the programme of~\cite{popl12} and develop a language with a new semantics and implementation, where orbit-finite sets are internally represented by set expressions over atoms.

\section{A basic functional language}
\label{sec:language}

To provide a functional language to construct and operate on definable sets over atoms, begin with a lambda calculus with a type $\mathsf{A}$ for atoms and a type $\mathsf{B}$ for boolean values, extended with a unary type constructor $\S$ that cannot be applied to values of function types. Thus types are defined by the following grammar:
\begin{align*}
	\tau &::= \mathsf{A} \mid \mathsf{B} \mid \S\tau \\
	\alpha,\beta &::= \tau \mid \alpha\to\beta
\end{align*}
The intuition is that values of type $\S\tau$ are (definable) sets of values of type $\tau$. This excludes function types, as one expects set elements to be equipped with a computable equality operation.

Terms of the core language are defined by the grammar:
\[
	M ::= C \mid x \mid \lambda x.M \mid MM 
\]
with the usual typing relation of lambda calculus, where $C$ comes from the following set of typed constants:
\begin{align*}
	{\tt empty} &: \S\tau  && \text{(the empty set)}\\
	{\tt atoms} &: \S\mathsf{A} && \text{(the set of all atoms)}\\
	{\tt insert} &: \tau\to \S\tau \to \S\tau && \text{(adds an element to a set)}\\
	{\tt map} &: (\tau_1\to\tau_2) \to \S\tau_1 \to \S\tau_2 && \text{(applies a function to every element)}\\
	{\tt sum} &: \S\S\tau \to \S\tau && \text{(union of a family of sets)}\\
	{\tt true}, {\tt false} &: \mathsf{B} && \text{(boolean values)} \\
	{\tt not} &: \mathsf{B} \to \mathsf{B} && \text{(logical negation)} \\
	{\tt and}, {\tt or} &: \mathsf{B} \to \mathsf{B} \to \mathsf{B} && \text{(conjunction and disjunction)}\\
	{\tt isEmpty} &: \S\tau \to \mathsf{B} && \text{(emptiness test)} \\
	{\tt if} &: \mathsf{B} \to \alpha \to \alpha \to \alpha && \text{(conditional)} \\
\intertext{We refrain from providing formal semantics for all these operations until the next section, but their
meaning should be intuitively clear as specified on the right above. Additionally, we include
some constants that depend on the signature of the underlying structure $\atoms$ of atoms. For equality atoms we take simply:}
	{\tt eq}_{\mathsf{A}} &: \mathsf{A}\to \mathsf{A} \to \mathsf{B} && \text{(equality relation on atoms)} \\
\intertext{and for ordered atoms, additionally:}
	{\tt leq} &: \mathsf{A} \to \mathsf{A} \to \mathsf{B} &&\text{(ordering relation on atoms)}.
\end{align*}
For other structures $\atoms$ this part of the language may change.

This core language can be extended with product types, integers, (mutually) recursive definitions, algebraic types and other features using standard techniques; we omit the details for brevity, noting only that the type metavariable $\tau$ should include all equality types. One can then define additional functions such as:
\begin{align*}
{\tt singleton} &: \tau\to\S\tau & {\tt singleton\ x} &= {\tt insert\ x\ empty} \\
{\tt filter} &: (\tau\to\mathsf{B})\to \S\tau \to \S\tau 
	& {\tt filter\ f\ s} &= {\tt sum}\ ({\tt map}\\&&&\hspace*{12pt} (\lambda{\tt x}.{\tt if}\ ({\tt f\ x})\ ({\tt singleton\ x})\ {\tt empty})\ {\tt s}) \\
{\tt exists} &: (\tau\to\mathsf{B})\to \S\tau \to \mathsf{B}
	& {\tt exists\ f\ s} &= {\tt not}\ ({\tt isEmpty}\ ({\tt filter\ f\ s})) \\
{\tt forall} &: (\tau\to\mathsf{B})\to \S\tau \to \mathsf{B}
	& {\tt forall\ f\ s} &= {\tt isEmpty}\ ({\tt filter}\ (\lambda x.{\tt not}\ ({\tt f\ x}))\ {\tt s}) \\
{\tt contains} &: \S\tau \to \tau \to \mathsf{B}
	& {\tt contains\ s\ x} &=  {\tt exists}\ ({\tt eq}\ {\tt x})\ {\tt s} \\
{\tt isSubsetOf} &: \S\tau \to \S\tau \to \mathsf{B}
	& {\tt isSubsetOf\ s\ t} &= {\tt forall}\ ({\tt contains\ t})\ {\tt s} \\
{\tt eq} &: \S\tau \to \S\tau \to \mathsf{B}
	& {\tt eq\ s\ t} &= {\tt and}\ ({\tt isSubsetOf\ s\ t})\\&&&\hspace*{33pt} ({\tt isSubsetOf\ t\ s}) \\
{\tt union} &: \S\tau \to \S\tau \to \S\tau 
	& {\tt union\ s\ t} &= {\tt sum}\ ({\tt insert\ s}\ ({\tt singleton\ t})) \\
{\tt intersection} &: \S\tau \to \S\tau \to \S\tau 
	& {\tt intersection\ s\ t} &= {\tt filter}\ ({\tt contains\ t})\ {\tt s}	
\end{align*}
and so on. In particular, for an equality type $\tau$, equality can be defined for the type $\S\tau$.

One can also construct sets definable by well-typed set expressions. For example,
\[
	{\tt atomPairs} = {\tt sum}\ ({\tt map}\ (\lambda{\tt x}.{\tt map}\ (\lambda{\tt y}.({\tt x},{\tt y}))\ {\tt atoms})\ {\tt atoms}) : \S(\mathsf{A},\mathsf{A})
\]
evaluates to the set of all pairs of atoms, and 
\[
	{\tt filter}\ (\lambda({\tt x},{\tt y}).{\tt not}({\tt eq}_{\mathsf{A}}\ {\tt x\ y}))\ {\tt atomPairs} : \S(\mathsf{A},\mathsf{A})
\]
to the set of all distinct pairs of atoms. 

In general, every set over atoms that is definable by a well-typed set expression is a value of some program. More formally, for every set expression $e:\tau$ with free variables $x_1,\ldots,x_k$ there is a term ${\tt set}_e:\mathsf{A}^k\to\tau$ in the programming language, that evaluates to a function from $\atoms^k$ that, when applied to arguments $a_1,\ldots,a_k$, returns $\sembr{e}_{[x_i\mapsto a_i]}$. This follows by induction on the structure of expressions. The only interesting case is 
\[
	e = \{e' : \phi \text{ for } x_{k+1},\ldots,x_m\} : \S\tau
\]
for some $e':\tau$ such that $FV(e'),FV(\phi)\subseteq\{x_1,\ldots,x_m\}$. It is easy to generalize the term {\tt atomPairs} above to a function
\[
	{\tt atomTuples}_{k,m} : \mathsf{A}^k \to\S\mathsf{A}^m
\]
that extends a given $k$-tuple of atoms to the set of all $m$-tuples that arise by putting arbitrary atoms on the remaining $m-k$ components. Then put
\[
	{\tt set}_e\ {\tt t} = {\tt map}\ {\tt set}_{e'}\ ({\tt filter}\ {\tt form}_{\phi}\ ({\tt atomTuples}_{k,m}\ {\tt t}))
\]
where ${\tt set}_{e'}$ exists by the inductive assumption, and ${\tt form}_{\phi}:\mathsf{A}^m\to\mathsf{B}$ is a term that encodes the first-order formula $\phi$. Such a term exists since $\atoms$ has quantifier elimination, and so without loss of generality we may assume that $\phi$ is quantifier-free.

\section{Logic-based semantics}\label{sec:semantics}

From the description of the language in Section~\ref{sec:language} it may not be clear how to implement operations postulated in it. For example, how to implement the function {\tt map} so that a function can be applied to every element of the infinite set of atoms in finite time? In this section we provide a (small-step) reduction semantics of the core functional language, that implements the set-theoretic intuitions provided in Section~\ref{sec:language}, yet is clearly computable.

The semantics is based on the following general ideas:
\begin{itemize}
\item Values of set types $\S\tau$ are represented not by enumerating their elements (that would be impossible, as usually they are infinite sets), but by set expressions as in Section~\ref{sec:sets-with-atoms}. 
\item Values of type $\mathsf{B}$ are not just boolean values; they are rather first-order formulas over a special kind of variables called {\em atom variables} that denote atoms.
\item Terms are evaluated in contexts that specify what relations hold between atom variables in them.
\item Sometimes a condition $\phi$ in a conditional expression ${\tt if}\ \phi\ M\ N$ is neither tautologically true nor false. In such cases it is not clear whether the conditional should evaluate to $M$ or $N$ and the choice is delayed for as long as possible. When delaying is not further possible, e.g. when $M$ and $N$ are atom variables, a {\em variant} is created that has value $M$ or $N$, formally depending on the value of $\phi$.
\end{itemize}

Formally, keeping the set of types as in Section~\ref{sec:language}, we extend the grammar of terms to:
\begin{align*}
	M &::= C\ \Big\vert\ x\ \Big\vert\ \lambda x.M\ \Big\vert\ MM\ \Big\vert\ a\ \Big\vert\ \phi\ \Big\vert\ \{M:\phi\text{ for }\sigma,\ldots,M:\phi\text{ for }\sigma\}\ \Big\vert\ M:\phi\vert\cdots\vert M:\phi
\end{align*}
where:
\begin{itemize}
\item $C$ ranges over the same set of typed constants as in Section~\ref{sec:language},
\item $a$ ranges over a fixed infinite set of {\em atom variables}, disjoint from the set of program variables such as $x$,
\item $\phi$ ranges over the set of first-order formulas (with quantifiers allowed) over the signature of $\atoms$ and over atom variables,
\item $\sigma$ ranges over finite sets of atom variables. We omit ``for $\sigma$'' if $\sigma$ is empty.
\end{itemize}
Note that the new terms are unavailable to the programmer and they shall appear only as final or intermediate values in the reduction semantics.

Atom variables in the sets $\sigma$ in set expressions are binding occurrences, just as the program variable $x$ is a binding occurrence in $\lambda x.M$. Terms are considered up to $\alpha$-equivalence, defined as expected. For example,
\[
	\{a:\neg(a=c)\text{ for }a\} \qquad \text{and}\qquad  \{b:\neg(b=c)\text{ for }b\}
\]
are $\alpha$-equivalent.

Expressions of the form
\[
	M_1:\phi_1\vert\cdots\vert M_n:\phi_n
\]
are called {\em variants}. They look syntactically similar to set expressions of the form $\{M_1:\phi_1,\ldots,M_n:\phi_n\}$, but their meaning is very different. A variant as above does not denote a set of values, but a {\em single} value whose identity cannot be determined at the moment and will be fixed depending on which one of the formulas $\phi_1$ to $\phi_n$ holds.

In addition to standard typing rules for the lambda calculus, the newly added terms are typed according to:
\begin{equation}\label{eq:typing2}
	a:\mathsf{A} 
	\qquad 
	\phi:\mathsf{B} 
	\qquad
	\dfrac{M_1:\tau \quad \cdots \quad M_n:\tau}
		{\{M_1:\phi_1\text{ for }\sigma_1,\ldots,M_n:\phi_n\text{ for }\sigma_n\}:\S\tau}
	\qquad
	\dfrac{M_1:\tau \quad \cdots \quad M_n:\tau}
		{(M_1:\phi_1\vert\cdots\vert M_n:\phi_n):\tau}
\end{equation}
relative to any typing context of free program variables in $M_1,\ldots,M_n$.

We define a small-step operational semantics where terms are evaluated in the context of a formula over atom variables. The basic semantic statements are of the form
\[
	\psi \vdash M\to N
\]
where $\psi$ is a formula and $M,N$ are program terms. Reduction rules are given in Fig.~\ref{fig:semantics}.

\begin{figure}
{\em $\beta$-reduction:}
\begin{equation}\label{rule:beta}
	\dfrac{\psi\vdash\ M\ \to\ M'}{\psi\ \vdash\ MN\ \to\ M'N}\qquad
	\dfrac{\psi\vdash\ N\ \to\ N'}{\psi\ \vdash\ MN\ \to\ MN'} \qquad
	\psi\ \vdash\ (\lambda x.M)\ N\ \to\ M[N/x]
\end{equation}
{\em Basic constants:}
\vspace*{-2ex}
\begin{gather}
\label{rule:const-start}
	\psi\vdash {\tt empty} \to \{\ \} \qquad \psi\vdash {\tt atoms} \to \{a:\top \text{ for } a\} \\[1ex]
	\psi\vdash {\tt insert}\ M\ \{M_1:\phi_1 \text{ for } \sigma_1,\ldots, M_n:\phi_n \text{ for } \sigma_n\}
			\to \{M:\top, M_1:\phi_1 \text{ for } \sigma_1,\ldots, M_n:\phi_n \text{ for } \sigma_n\} \\[1ex]
	\psi\vdash {\tt map}\ M\ \{M_1:\phi_1\text{ for }\sigma_1,\ldots,M_n:\phi_n\text{ for }\sigma_n\}
			\to \{MM_1:\phi_1\text{ for }\sigma_1,\ldots,MM_n:\phi_n\text{ for }\sigma_n\}\\[1ex]
\label{rule:sum}
	\begin{split}
	\psi\vdash {\tt sum}\ &\{\ldots,\{M_1:\phi_1\text{ for }\sigma_1,\ldots,M_n:\phi_n\text{ for }\sigma_n\}:\phi\text{ for }\sigma,\ldots\} \\
			&\to \{\ldots,M_1:\phi_1\land \phi\text{ for }\sigma_1\cup\sigma,\ldots,\ M_n:\phi_n\land\phi\text{ for }\sigma_n\cup\sigma,\ldots\}
	\end{split} \qquad \text{if } \sigma\cap\bigcup_{i=1}^n\sigma_i=\emptyset
		\\[1ex]
	\psi\vdash {\tt true} \to \top \qquad \psi\vdash {\tt false} \to \bot  \qquad \psi\vdash {\tt not}\ \phi \to \neg\phi\\[1ex] 
	\psi\vdash {\tt or}\ \phi_1\ \phi_2 \to \phi_1\lor\phi_2 \qquad \psi\vdash {\tt and}\ \phi_1\ \phi_2 \to \phi_1\land\phi_2 \\[1ex]
	\psi\vdash {\tt isEmpty}\ \{M_1:\phi_1\text{ for }\sigma_1,\dots,M_n:\phi_n\text{ for }\sigma_n\} 
			\to\ \bigwedge_{1\le i\le n} \underbrace{\forall{a_1}\forall a_2\cdots\forall a_k}_{\sigma_i=\{a_1,\ldots,a_k\}}. \neg \phi_i
\label{rule:const-end}
\end{gather}
{\em Conditional expressions:}
\vspace*{-2ex}
\begin{gather}
\label{rule:cond-start}	\dfrac{\atoms\models\psi\implies\phi}{\psi\vdash{\tt if}\ \phi\ M\ N \to M} 
	\qquad
	\dfrac{\atoms\models\psi\implies\neg\phi}{\psi\vdash{\tt if}\ \phi\ M\ N \to N}  \\[1ex]
\label{rule:cond-lambda}
	\dfrac{\atoms\not\models\psi\implies\phi \quad \atoms\not\models\psi\implies\neg\phi}
		{\psi\vdash{\tt if}\ \phi\ \lambda x.M\ \lambda x.N\ \to \lambda x.({\tt if}\ \phi\ M\ N)} \qquad
	\dfrac{\atoms\not\models\psi\implies\phi \quad \atoms\not\models\psi\implies\neg\phi}
	         {\psi\vdash{\tt if}\ \phi\ \phi_1\ \phi_2\to (\phi_1\land \phi)\lor(\phi_2\land\neg \phi)} \\[1ex]
\label{rule:cond-set}
	\dfrac{\atoms\not\models\psi\implies\phi \quad \atoms\not\models\psi\implies\neg\phi}
		{\splitdfrac{\psi\vdash{\tt if}\ \phi\ \{M_1:\phi_1\text{ for }\sigma_1,\ldots,M_n:\phi_n\text{ for }\sigma_n\}\ \{N_1:\theta_1\text{ for }\pi_1,\dots,N_k:\theta_k\text{ for }\pi_k\}}{\to\{M_1:\phi_1\land \phi\text{ for }\sigma_1,\ldots,M_n:\phi_n\land \phi\text{ for }\sigma_n,N_1:\theta_1\land\neg\phi\text{ for }\pi_1,\dots,N_k:\theta_k\land\neg \phi\text{ for }\pi_k\}}}  \\[1ex]
\label{rule:cond-atom}
	\dfrac{\atoms\not\models\psi\implies\phi \quad \atoms\not\models\psi\implies\neg\phi}
		{\psi\vdash{\tt if}\ \phi\ a\ b \to a:\phi | b:\neg \phi} \\[1ex]
	\dfrac{\atoms\not\models\psi\implies\phi \quad \atoms\not\models\psi\implies\neg\phi}		
		{\splitdfrac{\psi\vdash{\tt if}\ \phi\ (M_1:\phi_1|\cdots|M_n:\phi_n)\ (N_1:\theta_1|\cdots|N_k:\theta_k)}{\to M_1:\phi_1\land \phi|\cdots|M_n:\phi_n\land \phi|N_1:\theta_1\land\neg \phi|\dots|N_k:\theta_k\land\neg \phi}}  
\label{rule:cond-end}	
\end{gather}
{\em Set and variant reduction:}
\begin{gather}
\label{rule:under}
	\dfrac{\psi\land \phi \vdash\ M\to N}
		{\psi \vdash \{\ldots,M:\phi\textrm{ for }\sigma,\ldots\} \to \{\ldots,N:\phi\textrm{ for }\sigma,\ldots\}}
	\qquad
	\dfrac{\psi\land\phi \vdash M\to N}
		{\psi\vdash\cdots|M:\phi|\cdots\to \cdots|N:\phi|\cdots} \\[1ex]
\label{rule:variant-set}
	\psi \vdash \{\ldots,(M_1:\phi_1|\cdots|M_n:\phi_n):\phi\text{ for }\sigma,\ldots\}
		\to \{\ldots,M_1:\phi_1\land \phi\text{ for }\sigma,\ldots, M_n:\phi_n\land \phi\text{ for }\sigma,\ldots\}
\end{gather}
{\em Equality:}
\vspace*{-2ex}
\begin{gather}
\label{rule:eq}
	\psi\vdash {\tt eq}_{\mathsf{A}}\ (a_1:\phi_1|\dots|a_n:\phi_n)\ (b_1:\theta_1|\dots|b_m:\theta_m)
	\to\bigvee_{\genfrac{}{}{0pt}{}{1\le i\le n}{1\le j\le m}}(a_i=b_j \land \phi_i \land \theta_j)
\end{gather}
\caption{Reduction semantics}\label{fig:semantics}
\end{figure}
Rules~\eqref{rule:beta} provide the standard infrastructure of the lambda calculus. The notion of capture-avoiding substitution $M[N/x]$ works as usual taking into account the fact that atom variables in $\sigma$ bind in $\{M:\phi\text{ for }\sigma\}$. We do not commit to any particular reduction strategy allowing reductions both in functions and in their arguments.

Rules~\eqref{rule:const-start}--\eqref{rule:const-end} are mostly self-explanatory and they agree with the intuitive meaning of program constants as listed in Section~\ref{sec:language}. We only note that in rule~\eqref{rule:sum}, inner expressions $M_i:\phi_i\text{ for }\sigma_i$ may need to be $\alpha$-converted so that the side condition of the rule holds. Note also that the rule for {\tt atoms} in~\eqref{rule:const-start} is the only place where a new atom variable is created and that rule~\eqref{rule:const-end} may cause quantified first-order formulas to appear.

The conditional constant {\tt if} is evaluated in a special way and it deserves a separate section of the semantics. A premise $\atoms\models\psi\implies\phi$ means that the formula $\psi\implies\phi$ holds in $\atoms$ under every valuation of its free variables. If some valuation falsifies the formula, we write $\atoms\not\models\psi\implies\phi$. Rules~\eqref{rule:cond-start} apply where the value of the logical condition $\phi$ is determined by the ambient formula $\psi$. In such situations the condition $\phi$ behaves like a standard boolean value and the conditional expression is resolved as expected.

If the value of $\phi$ remains undetermined under the assumption of $\psi$, then both values to be chosen from must be combined in the result of the conditional expression. The course of action depends on the type of those values with the general idea to postpone the choice by pushing it down the structure of terms. If the two values are functions, in~\eqref{rule:cond-lambda} a new ``lazy'' function is created where the choice is postponed until the function argument is provided. If they are formulas or set expressions, 
rules~\eqref{rule:cond-lambda} and~\eqref{rule:cond-set} combine them in an expected way. The most interesting case is a choice between atom variables: in rule~\eqref{rule:cond-atom}, a variant is created. It may be seen as an ``ambiguous atom'' equal to $a$ or $b$ depending on the value of $\phi$. Formally, a separate rule~\eqref{rule:cond-end} for a choice between variants is required but it works as expected similarly to rule~\eqref{rule:cond-set}.

Notice that rule~\eqref{rule:cond-atom} is the only place where variants are created, and those variants are always built of atom variables. One may wonder why the typing rule for variants in~\eqref{eq:typing2} allowed arbitrary types $\tau$ instead of simply $\mathsf{A}$. This is in anticipation of other basic types added to the language such as integers or strings, excluded from the core language for brevity. For each such basic type, a rule corresponding to~\eqref{rule:cond-atom} would need to be added.

Variants tend to be short-lived intermediate values and they are dissolved as soon as they emerge as elements of set expressions. Rule~\eqref{rule:variant-set} shows how this is done. Rules~\eqref{rule:under} specify how reductions are done in the context of set expressions and variants; these rules show how ambient formulas $\psi$ are constructed.

Rule~\eqref{rule:eq} specifies the behaviour of the equality function used for equality atoms. This rule also applies to single atom variables which are here understood as degenerated variants $a:\top$. For ordered atoms the function {\tt leq} is specified analogously.

This reduction semantics has a few expected properties proved by standard arguments:
\begin{itemize}
\item {\em subject reduction} holds, i.e., the reduction relation preserves types,
\item the {\em Church-Rosser property} holds up to first-order formula equivalence, i.e., if $\phi\vdash M\to N$ and $\phi\vdash M\to N'$ then there exist terms $Q$ and $Q'$ such that $\phi\vdash N\to^*Q$ and $\phi\vdash N'\to^*Q'$, where $\to^*$ is the reflexive and transitive closure of $\to$, $Q$ and $Q'$ are equal up to replacing some first-order formulas with equivalent ones. This follows by a parallel reductions argument as described in~\cite{takahashi}.
\item {\em (weak) normalisation} holds, i.e., each term can be reduced to an irreducible value. This is proved by a standard type of argument~\cite{girard} assigning degrees to types of the language.
\end{itemize}

Obviously, normalization fails as soon as the core language is extended with recursion as non-terminating programs can then be written. Otherwise, the semantics can be routinely extended with product types and terms, integers, mutually recursive definitions, algebraic types, etc. This is illustrated by our implementation described in Section~\ref{sec:implementation}. Indeed, we do not implement the language from scratch; instead, we write a Haskell module to support features described here, allowing the programmer to use them in conjunction with the power of a full-fledged functional programming language.

\section{Hulls, supports and orbits}\label{sec:hulls}

In~\cite{popl12}, which is a direct predecessor to this paper, a different internal representation of infinite sets was used. To construct such sets a programming construction {\tt hull} was provided, which, given a finite list $C$ of atoms and a set of values $X$ of some type (possibly built of atoms), returned the closure of $X$ under all automorphisms of atoms that fix every element of $C$. For example, the expression
\begin{verbatim}
                                  hull [] {2}            
\end{verbatim}
evaluates to the set of all atoms, because every atom can be obtained from the atom ${\tt 2}$ by an application of an automorphism of $\atoms$ that fixes (which is a non-condition) every element of the empty list. Similarly,
\begin{verbatim}
                      hull [3] {2}            hull [2] {(2,5)}
\end{verbatim}
evaluate respectively to the set of atoms different from $3$, and to the set of pairs of atoms where the first element is $2$ and the second is different from $2$. If ordered atoms are considered, the expression
\begin{verbatim}
                                hull [] {(2,3)}
\end{verbatim} 
evaluates to the set of pairs where the second component is strictly greater than the first one. One could then manipulate sets constructed in this way using functions such as {\tt map} and {\tt sum}, so that, e.g., functions {\tt compose} and {\tt transitiveClosure} could be written more or less as in Section~\ref{sec:intro}. 
Internally, infinite sets were not represented by first-order formulas. Rather, the hull construction was used as a basic semantic construct in computed values of set types; see~\cite{popl12} for details.

The mechanism for representing infinite sets using hulls has a number of disadvantages. Most importantly, the size of the representation of an orbit-finite set is proportional to the number of its orbits. For example, the set of all triples of atoms is constructed by 
\begin{verbatim}
                hull [] {(1,1,1),(1,1,2),(1,2,1),(2,1,1),(1,2,3)}
\end{verbatim}
and in general the set of ordered $n$-tuples needs an internal representation of size exponential in $n$. This is rather inefficient and as a result in the prototype Haskell implementation of \nlambda{} from~\cite{popl12} only very rudimentary programs could be evaluated in reasonable time. Note that in our semantics the set of atom triples is represented internally by the more concise
\[
	\{(a_1,a_2,a_3): \top \text{ for } a_1,a_2,a_3\}.
\]
Another problem is that {\tt hull}-based definitions of sets require the use of constants that denote particular atoms, even if mathematical definitions of the same sets do not need to. For example, even though no concrete natural numbers are mentioned in a mathematical definition of triples of numbers, as many as three numbers are used in the {\tt hull}-based definition above. This is not a major problem when equality atoms are concerned, but with more sophisticated structures of atoms it would cause difficulties. For example, although the universal partial order~\cite{univposet} is a legal and well-behaved structure of atoms, no easy and natural representation of it is known and it is not clear how to denote its particular elements in a convenient way.

For these reasons, in this paper we replace the hull-based representation with the logic-based semantics from Section~\ref{sec:semantics}. One may even contemplate removing the {\tt hull} construction from the language available to the programmer, and indeed this is what we did for the core language in Sections~\ref{sec:language}--\ref{sec:semantics}. This is justified by the observation from Section~\ref{sec:language}, missed in~\cite{popl12}, that every definable set can be denoted by a program without {\tt hull}.
On the other hand, it is not clear how to define the hull function itself:
\[
	{\tt hull} : [\mathsf{A}] \to \S\tau \to \S\tau
\]
in the core language (extended with list types $[\alpha]$ in a standard way). As this function is sometimes useful to the programmer, we add it to the language along with a few other basic functions:
\begin{align*}
	{\tt groupAction} &: (\mathsf{A}\to\mathsf{A})\to \tau\to\tau &\qquad& \text{(renames free atoms in an argument)} \\
	{\tt supports} &: [\mathsf{A}] \to \tau \to \mathsf{B} &\qquad& \text{(checks if a list of atoms supports the argument)} \\
	{\tt support} &: \tau \to [\mathsf{A}] &\qquad& \text{(returns some finite support of the argument; efficient)} \\
	{\tt leastSupport} &: \tau \to [\mathsf{A}] &\qquad& \text{(returns the least support of the argument; less efficient)} \\
	{\tt setOrbit} &: \S\tau \to \tau \to \S\tau &\qquad& \text{(returns the orbit of an element in a set)} \\
	{\tt setOrbits} &: \S\tau \to \S\S\tau &\qquad& \text{(returns the (finite) set of orbits of a given set)}
\end{align*}
In~\cite{popl12}, most of these functions or their minor variations were derived from {\tt hull}. For example, one may write:
\begin{align*}
{\tt isSingleton} &: \S\tau\to\mathsf{B} & {\tt isSingleton\ s} &= 
	{\tt exists}\ (\lambda {\tt x}.{\tt forall}\ ({\tt eq\ x})\ {\tt s})\ {\tt s} \\
{\tt supports}&: [\mathsf{A}] \to \tau \to \mathsf{B} & {\tt supports\ c\ x} &=
	{\tt isSingleton}\ ({\tt hull\ c}\ ({\tt singleton\ x}))
\end{align*}
However, such definitions are rather inefficient. Here, we include {\tt groupAction} and {\tt support} as basic operations and define ${\tt hull}$ and other functions from them, which results in a more efficient implementation. 

\section{Implementation}\label{sec:implementation}

We implement \nlambda{} as a Haskell module (available from~\cite{nlambda}), which allows the programmer to use all benefits of a full-fledged functional programming language. The module introduces new types and functions operating on infinite structures and first-order formulas.

We shall now explain a few aspects of the implementation worth mentioning.

\subsubsection*{SMT solving}

In Fig.~\ref{fig:semantics}, rules~\eqref{rule:cond-start}--\eqref{rule:cond-end} involve premises of the form $\atoms\models\psi\implies\phi$, stating that a formula holds in the structure $\atoms$. Since we only consider $\omega$-categorical structures of atoms, one may equivalently ask whether $\psi\implies\phi$ follows from the axioms of the first-order theory of $\atoms$. This is an instance of the general satisfiability modulo theories (SMT) problem and there are software tools available that perform that task efficiently for a variety of first-order theories.

To determine whether a formula holds in $\atoms$ the interpreter of \nlambda{} calls the external Z3 solver~\cite{z3} via a system call. The implementation can be easily modified to connect to any other solver compatible with the SMT-LIB standard~\cite{smt-lib} instead. Currently two SMT-LIB logics are used: LIA (linear integer arithmetic, for equality atoms) and LRA (linear real arithmetic, for ordered atoms). Formula solving is a~pure function without side-effects, therefore it is invoked within the Haskell {\tt unsafePerformIO} function to avoid putting the {\tt IO} monad in types of all conditional statements in \nlambda. 

Experiments performed in our companion project LOIS \cite{lois} showed that SMT solvers in general and Z3 in particular do not deal well with quantified formulas that do not involve arithmetic. To improve performance before calling Z3, the interpreter eliminates all quantifiers from the formula to be checked. The quantifier elimination algorithm used for ordered atoms is based on the method of infinitesimals for linear real arithmetic proposed by Loos and Weispfenning \cite{loos-weispfenning} and adapted by Nipkow to dense linear order \cite{nipkow} (for equality atoms it is enough to use a simplified version of this algorithm). Roughly, this method involves replacing an existentially quantified formula by a disjunction of formulas where the bound variable is substituted by test points which include values arbitrarily close to either lower or upper bounds of the eliminated variable.

\subsubsection*{Conditionals}

From rules~\eqref{rule:cond-start}--\eqref{rule:cond-end} in Fig.~\ref{fig:semantics} it is clear that the conditional expression in \nlambda{} is substantially different from the standard Haskell ${\tt if...then...else...}$ construction, in that it must deal with conditions that cannot be resolved to {\tt true} or {\tt false}. Since {\tt if} is a Haskell keyword, a different name must be used for \nlambda{} conditionals; we choose
\begin{verbatim}
ite :: Conditional a => Formula -> a -> a -> a
\end{verbatim}
This function is implemented for all instances of the new {\tt Conditional} typeclass, which includes several basic types, the atom and formula types, list and function types. The function {\tt ite} first tries to determine the logical value of the condition formula with a SMT solver call; failing that, it calls a function {\tt cond} of the same type as {\tt ite} that is defined in a type-specific manner.

For example, the implementation of {\tt cond} for the formula type is:
\begin{verbatim}
instance Conditional Formula where
  cond f1 f2 f3 = (f1 /\ f2) \/ (not f1 /\ f3)
\end{verbatim}
For the function type it works in a lazy way:
\begin{verbatim}
instance Conditional b => Conditional (a -> b) where
  cond c f1 f2 = \x -> cond c (f1 x) (f2 x)
\end{verbatim}
These definitions correspond to rules~\eqref{rule:cond-lambda} in Fig.~\ref{fig:semantics}.

The result for the type of (definable) sets includes elements from both input sets but with appropriate formulas, according to rule~\eqref{rule:cond-set} in Fig.~\ref{fig:semantics}. In other collection types (lists, tuples, etc.), missing from  the core language of \nlambda{}, condition handling is passed to elements. The function for lists with the same lengths is coded as follows:
\begin{verbatim}
  cond c l1 l2 = zipWith (cond c) l1 l2
\end{verbatim}
One problem appears for an ambiguous condition on lists of different lengths. To simplify the implementation we decided to report an error in this case. However, operations on lists can be performed alternatively using the {{\tt Variants}} constructor.

\subsubsection*{Variants and contexts}

Of course some types (such as integer types) cannot cope with an ambiguous condition in any other way than to somehow return both values. For such types a special type constructor {\tt Variants} is provided; values of type {\tt Variants a} are lists of values of type {\tt a} coupled with formulas. It comes with its counterpart of {\tt ite} function, defined for any type {\tt a}:
\begin{verbatim}
iteV :: Formula -> a -> a -> Variants a
\end{verbatim}
Thus one can implement conditional statements e.g.~for integers: {\tt iteV (eq a b) 1 2} will return a variant $1:a = b\ |\ 2:a \neq b$, akin to rule~\eqref{rule:cond-atom} in Fig.~\ref{fig:semantics}. The type of atoms {\tt Atom} itself is actually defined as the variant type of variable names. Every variant type is an instance of the class {\tt Conditional}.

However, not always all possible result variants of the program are desired. Sometimes the result is interesting only in a given context. In such cases the new class {\tt Contextual} is useful. A function
\begin{verbatim}
when :: Contextual a => Formula -> a -> a
\end{verbatim}
introduces a formula into the context of a computation. For example, expression
\begin{verbatim}
when (neq a b /\ neq b c /\ neq a c) size (fromList [a,b,c])
\end{verbatim}
will display only the result for distinct atoms. This corresponds to adding formulas to contexts in rules~\eqref{rule:under} in Fig.~\ref{fig:semantics}.

\subsubsection*{Nominal types}

The basic type class in \nlambda{} is {\tt NominalType} corresponding to types ranged over by the $\tau$ metavariable in our core language. This class is required by several functions of the language and is important for three reasons:
\begin{itemize}
\item it provides an implementation of the equality predicate {\tt eq},
\item it has functions that operate on atom variables ({\tt mapVariables} and {\tt foldVariables}) and are used internally for resolving conflicts between atom variable names, and for collecting all or free atom variables that occur in a set expression, 
\item it helps split variant values into elements when inserting them to the set (to implement rule~\eqref{rule:variant-set} in Fig.~\ref{fig:semantics}).
\end{itemize}
To operate on a set of elements of a given type, the type has to be an instance of {\tt NominalType}.  Additionally, all instances of this class must be instances of the standard Haskell class {\tt Ord}. This is to improve performance.

\subsubsection*{Set types}

The {\tt Set} type constructor is an implementation of both infinite and finite sets. Generally, it is an alternative to the standard {\tt Data.Set} module with most features that can be found there. 
These include core functions of \nlambda{} such {\tt map}, {\tt filter} and {\tt sum} and functions defined from them as in Section~\ref{sec:language}. One can find auxiliary functions to deal with pairs, triples or in general tuples and lists of set elements. 

Notable omissions among functions provided by {\tt Data.Set} are those that rely on an ordering of set elements, such as {\tt elemAt}, {\tt toList} but also {\tt foldl} and {\tt foldr}. There seems to be no meaningful way to interpret these functions on infinite, definable sets.

One additional function that {\em is} provided calculates the size of a set:
\begin{verbatim}
size :: NominalType a => Set a -> Variants Int
\end{verbatim}
Certainly one can expect the answer in finite time only for finite sets. This function for consecutive natural numbers tries to find a list of distinct elements with a given length. This procedure is rather inefficient for large sets and does not terminate for infinite ones.

\subsubsection*{Hulls, supports and orbits}

As mentioned in Section~\ref{sec:hulls} and as will become apparent in Section~\ref{sec:examples} sometimes it is useful to the programmer to be able to operate on orbits of definable sets. For this purpose, functions listed in Section~\ref{sec:hulls} have been added to the language. The implementation of all these functions is derived from two basic ones:
\begin{verbatim}
support :: NominalType a => a -> [Atom]
groupAction :: NominalType a => (Atom -> Atom) -> a -> a
\end{verbatim}
The first returns a list of free atom variables in the argument (this list also serves as a support of it), the second applies a function to all free atom variables. Both functions invoke functions {\tt foldVariables} and {\tt mapVariables} that must be provided in instances of the {\tt NominalType} class.

Based on {\tt support} and {\tt groupAction} we implement the function
\begin{verbatim}
orbit :: NominalType a => [Atom] -> a -> Set a
\end{verbatim}
which computes the orbit of an element $e$ under the action of all automorphisms of $\atoms$ that fix all elements of a given support $[a_1,\dots,a_n]$. This function computes the list of free atoms $[b_1,\dots,b_k]$ in $e$, and filters all lists of atoms of length $k$:
\begin{equation*}
\{[x_1,\dots,x_k] :\textrm{ for }x_1,\dots,x_k \in\mathbb{A}\}
\end{equation*}
to obtain only these in the same orbit as $[b_1,\dots,b_k]$. To this end, a conjunction formula is built as follows:
\begin{equation*}
\mathop{\bigwedge}\limits_{\genfrac{}{}{0pt}{}{1\le i,j\le k}{i\neq j}} r(x_i,x_j)\iff r(b_i,b_j)\ \ \wedge\ \mathop{\bigwedge}\limits_{\genfrac{}{}{0pt}{}{1\le i\le k}{1\le j\le n}} r(x_i,a_j)\iff r(b_i,a_j)
\end{equation*}
for every relation $r$ in the signature of $\atoms$. (For equality atoms, it is just the equality relation.)
In the last step, the filtered set of lists is mapped with a function that replaces every atom $b_i$ in the element $e$ by $x_i$ for $1\le i \le k$.

Using {\tt orbit} an implementation of {\tt hull} and other functions listed in Section~\ref{sec:hulls} is now easy, for example:
\begin{verbatim}
hull :: NominalType a => [Atom] -> Set a -> Set a
hull supp = sum . map (orbit supp)
\end{verbatim}

\section{Examples}\label{sec:examples}

We demonstrate the potential and limitations of \nlambda{} on two simple examples: computing transitive closures of relations and graph $k$-colorability. Although both examples can be implemented in \nlambda, they are rather different. In the former one, standard Haskell code for calculating transitive closures of finite relations can be reused almost verbatim for the first-order definable case, sparing the programmer from considerations regarding finite vs.~infinite sets. In the latter example, standard Haskell code for finding $k$-colorings in finite graphs does not transport to the infinite setting. Instead, one partitions a given graph into its orbits, and looks for an {\em equivariant} coloring, where all nodes in the same orbit get the same color. Both the program and the proof of its correctness depend on the programmer's knowledge of first-order definable sets and their mathematical theory.

\subsubsection*{Transitive closures and cycles}

We begin by recalling the example presented in Section~\ref{sec:intro}. To compute the composition of two relations one can define a function {\tt compose} as follows: 

\begin{verbatim}
compose :: (NominalType a, NominalType b, NominalType c) => 
           Set (a,b) -> Set (b,c) -> Set (a,c)
compose r s = sum (map (\(a,b) -> 
                          map (\(_,c) -> (a,c)) 
                              (filter (eq b . fst) s)) 
                       r)
\end{verbatim}

This function can be written down more concisely, using some auxiliary functions. In \nlambda{} we provide some functions similar to the standard Haskell {\tt zip} and {\tt zipWith}:
\begin{verbatim}
pairs :: (NominalType a, NominalType b) => Set a -> Set b -> Set (a, b)
pairsWith :: (NominalType a, NominalType b, NominalType c) => 
             (a -> b -> c) -> Set a -> Set b -> Set c
\end{verbatim}

There are also functions that help filtering pairs:
\begin{verbatim}
pairsWithFilter :: (NominalType c, NominalType b, NominalType a) =>
     (a -> b -> NominalMaybe c) -> Set a -> Set b -> Set c
maybeIf :: Ord a => Formula -> a -> NominalMaybe a
\end{verbatim}

Using these one can implement {\tt compose} in a single line:
\begin{verbatim}
compose r s = pairsWithFilter (\(a, b) (c, d) -> maybeIf (eq b c) (a, d)) r s
\end{verbatim}

Now, one can code a function {\tt transitiveClosure} computing the transitive closure of a given relation:

\begin{verbatim}
transitiveClosure :: NominalType a => Set (a,a) -> Set (a,a)
transitiveClosure r = let r' = union r (compose r r)
                      in ite (eq r r') r (transitiveClosure r')
\end{verbatim}

It should be noted that the implementation of {\tt compose} and {\tt transitiveClosure} is similar to the finite version with only two differences: {\tt eq} instead of {\tt (==)} and {\tt ite} instead of an {\tt if...then...else...} statement.

Consider a datatype that describes directed graphs with vertices of any type and edges represented as pairs of vertices:
\begin{verbatim}
data Graph a = Graph {vertices :: Set a, edges :: Set (a,a)}
\end{verbatim}
To check whether a graph has a cycle one could use the function {{\tt transitiveClosure}} in the following way:
\begin{verbatim}
hasCycle :: NominalType a => Graph a -> Formula
hasCycle (Graph vs es) = exists (uncurry eq) (transitiveClosure es)
\end{verbatim}
When only odd-length cycles are requested, one could define a function {\tt hasOddLengthCycle} as presented below:

\begin{verbatim}
hasOddLengthCycle :: NominalType a => Graph a -> Formula
hasOddLengthCycle (Graph vs es) = intersect (map swap es) 
                                            (transitiveClosure (compose es es))
\end{verbatim}
where {\tt (transitiveClosure (compose es es))} returns the set of all pairs of vertices connected with even-length paths. If some pair of vertices from this set is also connected with an edge from the original graph, it means that there is an odd-length cycle.

Note how the above fragments of code are essentially the same as ones that would be used for computing transitive closures or cycle finding on finite graphs.

\subsubsection*{Graph coloring}

Recall that a graph coloring is a valuation of its nodes such that no two adjacent vertices share the same value. The verification whether a given function is a valid coloring looks as follows:

\begin{verbatim}
isColoringOf :: (NominalType a,NominalType b) => (a -> b) -> Graph a -> Formula
isColoringOf c g = forAll (\(v1,v2) -> c v1 `neq` c v2) (edges g)
\end{verbatim}

A $k$-coloring is a graph coloring with $k$ colors. In order to check whether a graph is $k$-colorable in the finite setting, one could generate all $k$-partitions of a set of $n$ vertices: 
\begin{verbatim}
partitions :: Int -> Int -> Set [Int]
partitions n 1 = singleton (replicate n 0)
partitions n k | k < 1 || n < k = empty
partitions n k | n == k = singleton [0..n-1]
partitions n k = union (map (k-1:) $ partitions (n-1) (k-1)) 
                       (pairsWith (:) (fromList [0..k-1]) (partitions (n-1) k))
\end{verbatim}
For example, ({\tt partitions} 3 2) evaluates to a set of three partitions: {\tt \{[0,0,1], [1,0,0], [1,0,1]\}}. For each such partition one could examine if the valuation that arises from it is a valid coloring. 

In the world of definable sets the situation is much more complicated. One cannot enumerate and collect all partitions because the set of partitions of a definable set might not be first-order definable or even countable. Indeed, at first sight it is not clear that colorability of definable graphs is a decidable problem. 
For example, consider the undirected graph:
\begin{align}\label{eq:satangraph}
\begin{split}
\mathtt{Graph}\ \{&\mathtt{vertices} = \{(a_1,a_2) : a_1 \neq a_2\ \textrm{for}\ a_1,a_2 \in \mathbb{A}\},\\
&\mathtt{ edges} = \{\{(a_1,a_2),(a_2,a_3)\} : a_1 \neq a_2\land a_1\neq a_3\land a_2\neq a_3\ \textrm{for}\ a_1,a_2,a_3 \in \mathbb{A}\}\}
\end{split}
\end{align}
This graph, used as an example in~\cite{KKOT15}, is not 3-colorable. However, its smallest finite non-3-colorable graph has as many as 10 vertices and 20 edges. One may try to check larger and larger finite subgraphs of a given definable graph and check their colorability using the standard code above, but it is not clear when one can stop and declare the entire graph colorable.

One may make some additional assumptions, for example consider only {\em equivariant} colorings, where nodes in the same orbit must get the same color. (For example, the graph in~\eqref{eq:satangraph} has no equivariant colorings, as it only has one orbit of vertices and it has edges.) The problem then reduces to coloring the finite set of orbits.
For a given list of orbits and a list of its partitions one can create a coloring function that determines which orbit contains a given element and returns the color assigned to such an orbit.

\begin{verbatim}
coloring :: NominalType a => [Set a] -> [Int] -> a -> Variants Int
coloring [] [] _ = variant 0
coloring (o:os) (p:ps) a = ite (member a o) (variant p) (coloring os ps a)
\end{verbatim}

Then it remains to check whether a coloring function created by a partition of orbits is a proper coloring of the graph. This can be implemented as follows:

\begin{verbatim}
hasEquivariantColoring :: NominalType a => Graph a -> Int -> Formula
hasEquivariantColoring g k = member true $
    pairsWith (\os ps -> (coloring os ps) `isColoringOf` g) 
              (replicateSet n orbits) 
              (partitions n k)
    where orbits = setOrbits (vertices g)
          n = maxSize orbits
\end{verbatim}
where {\tt replicateSet :: NominalType a => Int -> Set a -> Set [a]}
returns the set of lists with a given length and elements from a set.

This solves the problem of finding equivariant colorings of definable graphs. As it turns out it solves the problem of general $k$-colorability as well: in~\cite{KKOT15}, it was proved that {\em over ordered atoms} a definable graph has a $k$-coloring if and only it has an equivariant one. That result relies on deep theorems in topological dynamics. 
As we can see, the programmer needs to know the mathematics of first-order definable structures not only to write the program for $k$-colorability, but even more so to prove its correctness.

It is worth noting that the problem of finding an equivariant $k$-coloring may have different solutions depending on the structure of atoms. For example, the graph:
\begin{align*}
{\tt g\ =\ }\mathtt{Graph}\ \{&\mathtt{vertices} = \{(a_1,a_2) : a_1 \neq a_2\ \textrm{for}\ a_1,a_2 \in \mathbb{A}\},\\
&\mathtt{ edges} = \{((a_1,a_2),(a_2,a_1)) : a_1 \neq a_2\ \textrm{for}\ a_1,a_2 \in \mathbb{A}\}\}
\end{align*}
does not have an equivariant $2$-coloring when equality atoms are considered. But for ordered atoms, a function {\tt (uncurry lt)} with type: {\tt (Atom, Atom) -> Formula} is a correct coloring. So for these two structures of atoms the expression {\tt (hasEquivariantColoring g 2)} will evaluate to {\tt false} and {\tt true} respectively.

Note that $2$-colorings can be looked for in a way very similar to the one used for finite graphs; indeed, a graph is $2$-colorable if and only if it has no cycle of odd length, and an \nlambda{} program to check that was shown above. The expression {\tt (hasOddLengthCycle g)} will evaluate to {\tt false} both over equality and ordered atoms, indicating that a $2$-coloring (not necessarily equivariant) of ${\tt g}$ exists.

These are only selected examples of programs in \nlambda. We have also solved problems such as reachability, finding weakly or strongly connected components in graphs, the emptiness problem of automata~\cite{lmcs14} and a minimization algorithm of automata. None of these require the programmer to explicitly use orbits and other structure of definable sets. However, as the example of graph $k$-coloring (for $k>2$) shows, certain problems do seem to require that. We do not understand precisely what it means for a problem to ``require the use of orbits'' or where the division lies between problems that do or do not. A possible connection to descriptive complexity theory and the celebrated ``quest for PTIME logic''~\cite{grohe-quest} could be imagined but this is left for future work.

\bibliographystyle{eptcs}
\bibliography{main}

\end{document}